\documentclass[preprint, 3p]{elsarticle}
\pdfoutput = 1
\usepackage{lipsum}
\makeatletter
\def\ps@pprintTitle{%
\let\@oddhead\@empty
\let\@evenhead\@empty
\def\@oddfoot{}%
\let\@evenfoot\@oddfoot}
\makeatother

\usepackage{algorithm}
\usepackage{algpseudocode}
\usepackage{amsmath}
\usepackage{amssymb}
\usepackage{amsthm}
\usepackage{amscd}
\usepackage{appendix}
\usepackage{bbm}
\usepackage{braket}
\usepackage{bookmark}
\usepackage{booktabs}
\usepackage{caption}
\usepackage{cases}
\usepackage{enumitem}
\usepackage[T1]{fontenc} 
\usepackage{float}
\usepackage{lineno,hyperref}
\modulolinenumbers[5]
\usepackage{indentfirst}
\usepackage[utf8]{inputenc}
\usepackage[autostyle,italian=guillemets]{csquotes}
\usepackage{graphicx}
\graphicspath{{Images/}}
\DeclareGraphicsExtensions{{.png},{.pdf},{.jpg}}
\graphicspath{{./}}
\usepackage{lmodern}
\usepackage{lipsum} 
\usepackage{listings}
\usepackage{yhmath}
\usepackage{mathtools}
\usepackage{mathrsfs}

\usepackage{microtype} 
\usepackage{multirow}
\usepackage{setspace}
\usepackage{standalone}
\usepackage{subcaption}
\usepackage{tabularx}
\usepackage[table]{xcolor}
\usepackage{wasysym}
\usepackage{wrapfig}
\usepackage{url} 
\usepackage{wrapfig}
\usepackage{lscape}
\usepackage{rotating}
\usepackage{siunitx}
\usepackage[textsize=footnotesize]{todonotes}

\usepackage[overload]{empheq}
\usepackage{cleveref} 
\usepackage[acronym]{glossaries}
\makeglossaries

\newacronym{eam}{EAM}{electro-anatomical mapping}
\newacronym{ecg}{ECG}{electro-cardiogram}
\newacronym{egm}{EGM}{electrogram}
\newacronym{csp}{CSP}{conduction system pacing}
\newacronym{cdt}{CDT}{cardiac digital twin}
\newacronym{cep}{CEP}{cardiac electrophysiology}
\newacronym{isct}{ISCT}{\emph{in silico} clinical trial}
\newacronym{bspm}{BSPMs}{body surface potential maps}
\newacronym{utc}{UTCs}{universal torso coordinates}
\newacronym{vt}{VT}{ventricular tachycardia}
\newacronym{af}{AF}{atrial fibrillation}
\newacronym{lat}{LAT}{local activation time}
\newacronym{hps}{HPS}{His-Purkinje system}
\newacronym{rmse}{rmse}{relative root mean square error}
\newacronym{cc}{CC}{correlation coefficient}
\newacronym{ep}{EP}{electrophysiology}
\newacronym{ab}{AB}{apico-basal}
\newacronym{si}{SI}{superior-inferior}
\newacronym{pa}{PA}{posterior-anterior}
\newacronym{rl}{RL}{right-left}
\newacronym{pca}{PCA}{principal component analysis}
\newacronym{rotab}{ROT-AB}{rotation apico-basal}
\newacronym{rotpa}{ROT-PA}{rotation posterior-anterior}
\newacronym{rotrl}{ROT-RL}{rotation right-left}
\newacronym{trsi}{TR-SI}{translation superior-inferior}
\newacronym{trpa}{TR-PA}{translation posterior-anterior}
\newacronym{trrl}{TR-RL}{translation right-left}

\newdefinition{rmk}{Remark}
\newproof{prf}{Proof}

\begin{document}
	
\begin{frontmatter}
\title{Quantifying variabilities in cardiac digital twin models of the electrocardiogram}

\author[1]{Elena Zappon}
\ead{elena.zappon@medunigraz.com}
\author[1]{Matthias A.F. Gsell}
\ead{matthias.gsell@medunigraz.at}
\author[1,2,3]{Karli Gillette}
\ead{karli.gillette@medunigraz.at}
\author[1,4]{Gernot Plank\corref{cor1}}
\ead{gernot.plank@medunigraz.at}

\cortext[cor1]{Corresponding author}
		
\address[1]{Division of Biophysics and Medical Physics, Gottfried Schatz Research Center, Medical University of Graz, Graz, Austria}

\address[2]{Scientific Computing and Imaging Institute, University of Utah, SLC, UT, USA}
\address[3]{Department of Biomedical Engineering, University of Utah, SLC, UT, USA}

\address[4]{BioTechMed-Graz, Graz, Austria}

%

\begin{abstract}
	\Glspl{cdt} of human cardiac \gls{ep} are digital replicas of patient hearts 
	that match like-for-like clinical observations. 
	The \gls{ecg}, as the most prevalent non-invasive observation of cardiac electrophysiology, is considered an ideal target for \gls{cdt} calibration. Recent advanced \gls{cdt} calibration methods have demonstrated their ability to minimize discrepancies between simulated and measured \gls{ecg} signals, effectively replicating all key morphological features relevant to diagnostics. However, due to the inherent nature of clinical data acquisition and \gls{cdt} model generation pipelines, discrepancies inevitably arise between the real physical electrophysiology in a patient and the simulated virtual electrophysiology in a \gls{cdt}.

    In this study, we aim to qualitatively and quantitatively analyze the impact of these uncertainties on \gls{ecg} morphology and diagnostic markers. We analyze residual beat-to-beat variability in \gls{ecg} recordings obtained from healthy subjects and patients. Using a biophysically detailed and anatomically accurate computational model of whole-heart electrophysiology combined with a detailed torso model calibrated to closely replicate measured \gls{ecg} signals, we vary anatomical factors (heart location, orientation, size), heterogeneity in electrical conductivities in the heart and torso, and electrode placements across \gls{ecg} leads to assess their qualitative impact on \gls{ecg} morphology.
    
    Our study demonstrates that diagnostically relevant \gls{ecg} features and overall morphology appear relatively robust against the investigated uncertainties. This resilience is consistent with the narrow distribution of \Gls{ecg} due to residual beat-to-beat variability observed in both healthy subjects and patients.

\end{abstract}

\end{frontmatter}

\section{Introduction}
Computational modeling of cardiac \gls{ep} is an established important research tool for analyzing experimental or clinical data \cite{Cluitmans2023,Monaci2022,Heijman2021,aronis2019understanding,KABRA2022263}, and is now increasingly considered in industrial applications such as medical device design \cite{swenson2020:_atp,corral2020digital}, 
as well as in clinical applications for diagnosis, stratification, and therapy planning \cite{peirlinck2021precision,niederer2019computational,cartoski2019computational,arevalo2016arrhythmia}.
Unlike in basic research, where generic representations of cardiac anatomy and \gls{ep} are used to gain generic mechanistic insights, a more specific individualized modeling approach is required in industrial and clinical applications \cite{trayanova2023computational,lopez2019personalized,gray2018patient,viceconti2016silico}. 
There, models must be calibrated to cover the variability of a given patient population, 
in the form of a virtual cohort \cite{niederer2019computational},
or, even to represent anatomy and \gls{ep} of individual patients. Such high-fidelity digital replicas of patient hearts, derived from clinical data and calibrated to match like-for-like clinical observations, 
are often referred to as patient-specific models or \glspl{cdt} \cite{corral2020digital,trayanova2024up,gsell2023towards,jung2022integrated}.
 
Owing to its broad clinical availability and non-invasive nature, \gls{ecg} recordings appear to be a most natural choice as a target for model calibration \cite{grandits2023digital,qiao2023dual,camps2023digital}.
Advanced \gls{cdt} calibration methods have been developed that are, in principle, able to minimize the discrepancy between simulated and measured \gls{ecg} \cite{grandits2023digital,gillette2021:_framework,CAMPS2024103108}, 
and have been shown to replicate all morphological key features relevant for diagnostics \cite{gillette2021:_framework,gillette2023medalcare}.
 These methods have demonstrated the ability to replicate all key morphological features relevant for diagnostics \cite{gillette2021:_framework,gillette2023medalcare}. However, inconsistencies inevitably arise between the real physiological electrical activity in a patient and the simulated virtual electrical activity in a \gls{cdt}, due to the nature of clinical data acquisition and \gls{cdt} model generation pipelines.
Major sources of these inconsistencies include: i) the fundamental issue of non-uniqueness in the models used to solve the forward problem of electrocardiography -- different activation and repolarization sequences can produce identical \gls{ecg} patterns \cite{xanthis2007inverse}; ii) residual variability in heartbeat, leading to beat-to-beat variations in \gls{ecg} morphology \cite{hasan2012relation,FRLJAK2003267,APPEL19891139}, which limits the fidelity of a \gls{cdt} calibrated using a single \gls{ecg}; and iii) significant observational uncertainties due to technical limitations in anatomical imaging and \gls{ecg} recording, which hinder accurate simultaneous measurement of all model parameters contributing to \gls{ecg} generation \cite{keall2006management,hawkes2005tissue}. 
Therefore, all factors involved in anatomical and electrophysiological modeling must be considered as uncertainty, and the measured \gls{ecg} itself as one signal of an envelope of \gls{ecg}s.

Residual \gls{ecg} variability arises from physiological sources, such as beat-to-beat variations in the activation sequence \cite{Monfredi2013,Zaniboni2000}, and anatomical variations of the heart due to factors like breathing \cite{holst2018respiratory,claessen2014interaction,Shechter2004,lendrum1979respiratory,JAGSI2007253,Shechter2004} and body posture \cite{rodeheffer1986postural,rapaport1966effect}, which alter the shape, position, and orientation of the heart relative to \gls{ecg} recording locations on the torso \cite{MACLEOD2000229,hoekema2001geometrical,van2000geometrical}. In addition to anatomical factors, there are significant uncertainties in the torso structure due to electrical heterogeneities \cite{ramanathan2001electrocardiographic,benjamin1950electrical}, and related model parameters that cannot be directly measured \emph{in vivo}, only estimated. Furthermore, despite standardization of \gls{ecg} lead placement, operator variability in electrode positioning remains non-negligible \cite{medani2018accuracy,kania2014effect,lateef2003vertical,Hill1987561,kerwin1960method}. If variability in any of these factors translates into marked variability in \gls{ecg} morphology, it could render \gls{ecg}-based calibration ambiguous and unreliable.

In this study, we aim to qualitatively and quantitatively analyze the impact of the latter two sources of uncertainties on \gls{ecg} morphology and diagnostic markers, and to investigate their role in the \gls{ecg}-based calibration of \gls{cdt}. To this end, we investigate residual beat-to-beat variability in \gls{ecg} recordings obtained from healthy subjects and patients undergoing treatment for atrial fibrillation or ventricular tachycardia ablation therapy. As a computational reference, a biophysically detailed anatomically accurate computational whole heart-torso model of one of the analyzed healthy subjects is calibrated to replicate the measured mean \gls{ecg} with high fidelity. 
Keeping constant the electrical activation and repolarization sequences, we vary the position, orientation, and size of the heart, heterogeneity in electrical conductivities in the torso, 
and electrode placement in individual \gls{ecg} leads, to compare their impact on the \gls{ecg}.  Our findings indicate that diagnostically relevant \gls{ecg} features and overall morphology remain relatively robust against the investigated uncertainties. This resilience aligns with the narrow distribution of \gls{ecg} due to residual beat-to-beat variability observed in both healthy subjects and patients. Therefore, our results suggest that an accurate inference of \gls{cdt} electrical activation sequences from the \gls{ecg} is feasible, and not impeded by residual beat-to-beat variability or inevitable model inconsistencies. 
\medskip

\section{Methods}


\subsection{Residual Variability}
We analyzed beat-to-beat residual variability in the 12-lead \gls{ecg} for a cohort of 14 healthy subjects, 
and patients treated for \gls{vt} and \gls{af}. 
To provide a relative margin of uncertainty, over \SI{10}{\second} \glspl{ecg} were recorded. The \gls{ecg} data were then filtered with a 150 Hz low pass filter, a 50 Hz bandstop filter, and a high pass filter of 0.05 Hz to reduce noise. 
Individual beats were detected using a modified Pan Tompkins algorithm for R-wave detection \cite{bernd_porr_2023_7652725} and stored in a beat matrix.
For each matrix, the average \gls{ecg} beat was computed and plotted against all beats in the matrix. 


Normal sinus rhythm \glspl{ecg} were recorded from 18 \gls{af} and 17 \gls{vt} patients during ablation procedures
using an electro-anatomical mapping system (Carto XP).
Ablation procedures were carried out at the University Hospital of Graz, Graz, Austria, and included in the local ablation registry approved by the ethics committee of the Medical University of Graz (reference number 31-036 ex 18/19 for the \gls{vt} patients, and reference number 26-217 ex 13/14 for the \gls{af} patients). All patients gave written informed consent.
\Glspl{ecg} of \SI{2.5}{\second} were recorded using an electro-anatomical mapping system (Carto XP)
at each position visited by the mapping catheter. The \gls{ecg} corresponding to the beat selected by the mapping system for deriving an instance of local activation was chosen. Depending on the density of the constructed maps, hundreds to thousands of \glspl{ecg} were recorded, analyzed \cite{ARNOLD2024108299}, and stored per patient in a beat matrix.
As for the healthy subjects, an average \gls{ecg} was computed and plotted against the entire beat matrix.

\subsection{ECG modeling}
The reference baseline forward \gls{ecg} model has been described previously in great detail elsewhere \cite{gillette2021:_framework} with updates as detailed in \cite{gillette2022:_personalized} and \cite{gillette2021:_hps}.
Briefly, an anatomically accurate heart-torso model of the subject was generated from clinical magnetic resonance imaging \cite{crozier2016:_image_based}. 
Images of the heart were segmented using an automated tool \cite{payer2018:_multi_label}, and the segmentation was semi-automatically refined using \texttt{seg3D} \cite{SCI:Seg3D}. The segmented patient-specific anatomy included a whole heart embedded in a torso with cardiac blood pools, lungs, bones, liver, fat, and skin labeled as different volumes. Computational meshes were generated from segmentation labels  \cite{prassl2009:_meshing} using \cite{neic2020:_meshtool}, at an average resolution of \SI{1224}{\micro\meter} in the heart, and a coarser resolution in other tissues and torso, 
with an average resolution of \SI{3444}{\micro\meter} on the torso surface. Myocardial fibers were implemented based on rule-based algorithms within the atria \cite{roney2021constructing} and ventricles \cite{Bayer2012}. Universal coordinates were computed within the atria \cite{roney2019universal} and the ventricles \cite{gillette2021:_framework,bayer2018universal}. 

The model was calibrated to faithfully replicate the measured \gls{ecg} from this subject across all 12 leads. The atria were assigned generic electrophysiology that gave a realistic P-wave as detailed with activation stemming from a sino-atrial node on the right atrial roof. An atrio-ventricular node within the right atria was connected to a physiologically detailed His-Purkinje system that facilitated ventricular activation. Initially, a simplified ventricular conduction system was assumed to compromise 5 fascicles rooted in the endocardium -- 3 in the left ventricle, one in the right-ventricular septum, and one in the right ventricular moderator band, combined with a fast-conducting endocardial layer. Fascicular locations were varied through sampling to minimize the mismatch in QRS morphology in the \gls{ecg}. The optimized five-fascicle conduction system was subsequently replaced by a topologically realistic model of the \gls{hps} that produced the same actThe model was calibrated to faithfully replicate the measured \gls{ecg} from this subject across all 12 leads.ivation sequence and retained an equally good match in the \gls{ecg} \cite{gillette2021:_hps}. Action potential duration was spatially varied using a linear mapping with activation to obtain heterogeneous ventricular repolarization patterns satisfyingly matching the T-wave morphology \cite{gillette2022:_personalized}. All other electrophysiological parameters are described in further detail in \cite{gillette2022:_personalized}. 

\begin{figure}[!t]
	\centering
	\includegraphics[width=1\textwidth]{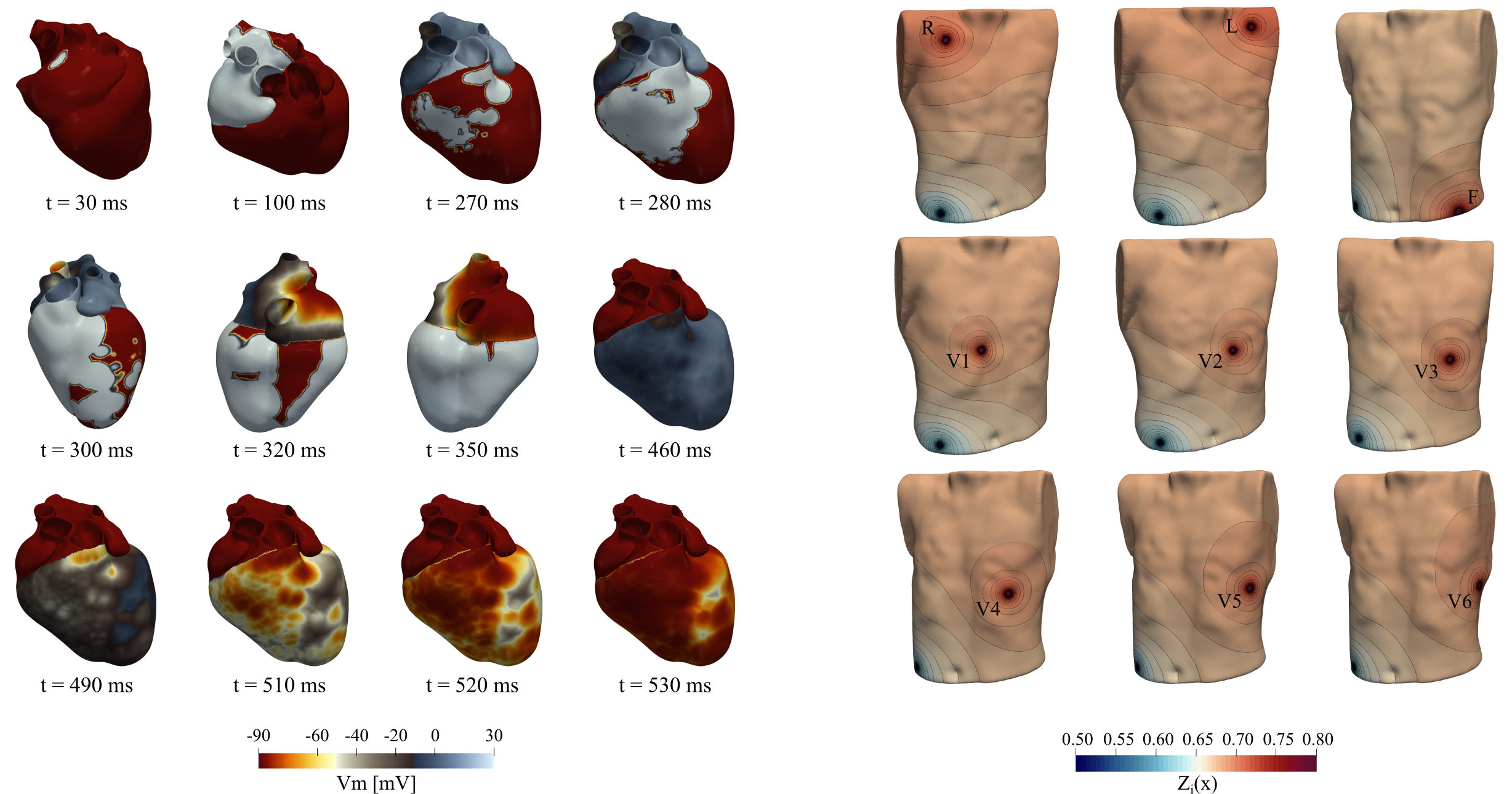}
	\caption{Left: Transmembrane potential $V_{\rm m}(\mathbf{x},t)$ for the reference simulation. Right: Lead field solutions $Z_i(\mathbf{x})$ considering the 9 electrodes position R, L, F, V1, V2, V3, V4, V5, and V6 to compute the 12-lead \gls{ecg}. The range of $Z_i(\mathbf{x})$ solutions, which is $[-1.0, 1.4]$, is restricted between $[0.5, 0.8]$ to improve the visualization.}	
	\label{fig:potential}
\end{figure}

The reaction-Eikonal model in monodomain mode was used to describe wavefront propagation and the associated electrical sources in the form of transmembrane voltages, $V_{\rm m}(\mathbf{x},t)$ (referred to Fig.\ref{fig:potential}, left), 
generating the \gls{ecg} \cite{neic2017:_reaction_eikonal}. 
For all \gls{ecg} recording locations, $\mathbf{x}_i$, the Lead field solutions, $Z_i(\mathbf{x})$ (referred to Fig.\ref{fig:potential}, right), 
have been computed and used to accurately compute extracellular potential differences, $V(t)$, 
between electrode locations corresponding to the \gls{ecg} \cite{potse2018scalable}.

\subsection{Cardiac anatomical variation}
The role of anatomical variation of the heart in position, orientation, and size, 
mediated due to breathing and body posture, upon the \gls{ecg} is investigated by defining a set of heart-torso anatomies while preserving the EP settings and simulation from the calibrated reference model. Starting from the reference cardiac and torso geometry, anatomical uncertainty was introduced by altering the heart's position, orientation, and size. The external shape of the torso and the electrode placements were kept constant for all anatomical models. 
To prevent intersections between the heart and other organs or tissues, a homogeneous torso model was used for this investigation.

\begin{figure}[!t]
	\centering
	\includegraphics[width=0.7\textheight]{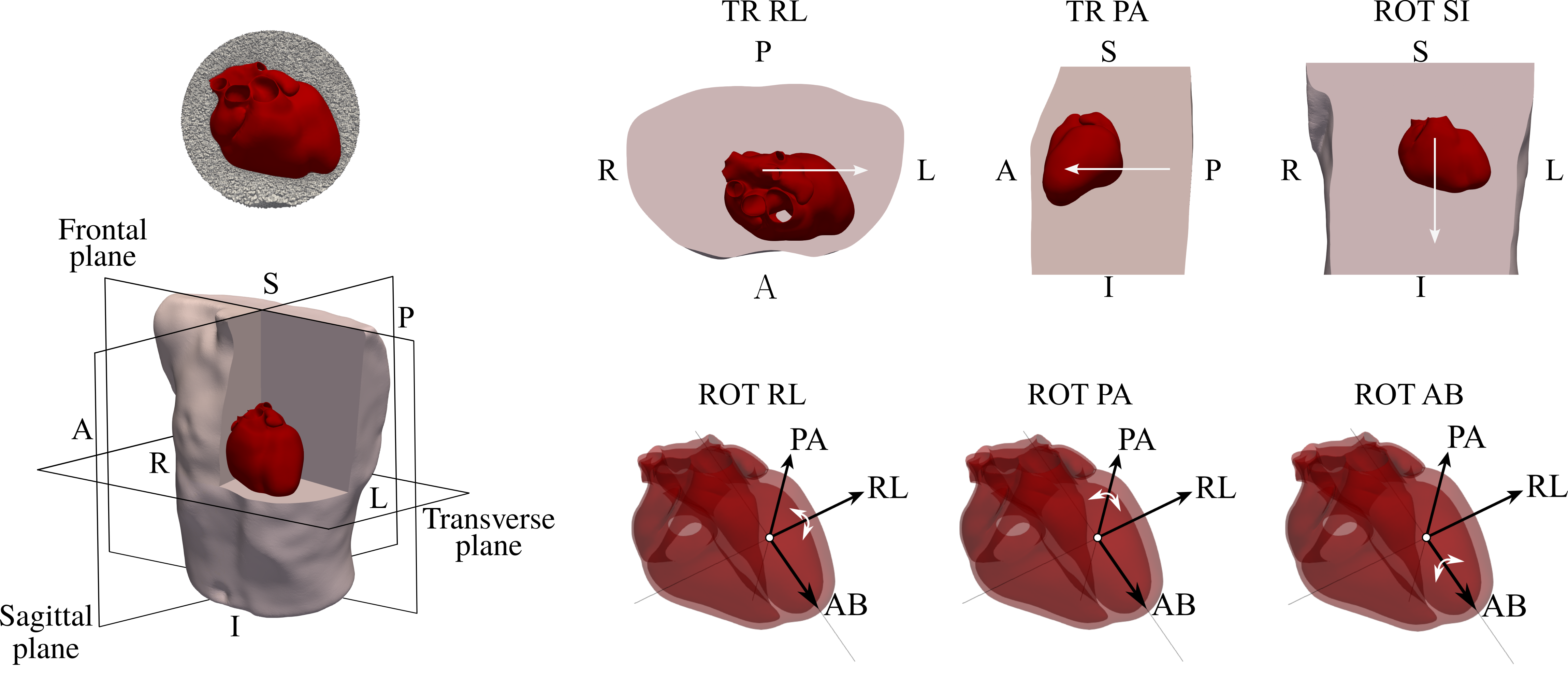}
	\caption{Top-left: cardiac model embedded in a spherical halo. For each investigated position of the heart, the halo is the only remeshed part of the torso. Bottom-left: schematic representation of the frontal, sagittal, and transverse plane of the torso. Top-right: schematic representations of the three directions of translation. Bottom-right: schematic representations of the three rotational axes.}	
	\label{fig:transl_rot}
\end{figure}

To circumvent the need for complete remeshing of the torso with each new cardiac configuration, we introduced a spherical halo surrounding the heart (referred to Fig.~\ref{fig:transl_rot}). 
The volumetric meshes extending from the surface of the torso to the halo, as well as the heart mesh itself, remained unchanged for all anatomical models, with only the smaller volume between the halo and each new heart position, orientation, and shape requiring remeshing. To streamline the generation of the anatomical set, an efficient and semi-automatic pipeline was implemented. 
 
Variations in position, orientation, and size of the heart were modeled by rigid translations, rotations, and scaling of the heart. 
To cover the full range of possible anatomical cahnges\cite{JAGSI2007253,Shechter2004}, 
translational directions and rotational axes were built following the work of Odille et al.  \cite{odille2017statistical}, 
where the authors present a cardiac reference system that aids the statistical investigation of the variation of the heart position in the human population. 
Three main translational directions, each with corresponding positive and negative vectors, were defined, along with three rotational axes and their positive and negative rotational angles (see Fig.~\ref{fig:transl_rot}). 
To ascertain the exploration of the entire space of anatomical variability of the heart, we employed translational, rotational, and scaling magnitudes well beyond the physiological ranges as detailed in the following.

For all changes in the heart geometry, 
the electrical source distribution $V_{\rm m}(\mathbf{x},t)$ over the entire myocardial volume was kept constant, while the lead field solutions $Z_i(\mathbf{x})$ were recomputed, thus minimizing any potential impact of spatial discretization upon depolarization and repolarization pattern on the \gls{ecg} prediction.

\emph{Position uncertainty} was accounted for by translating the heart within the torso along three orthogonal directions. These were determined using the cardiac center of mass as origin, and axes aligned perpendicular to the frontal, sagittal, and transverse planes of the torso (referred to Figure \ref{fig:transl_rot}). 
The three axes were labeled according to their anatomical orientation within the torso as follows:
perpendicular to the sagittal plane as \gls{rl}, 
to the frontal plane as \gls{pa}, and to the transverse plane as \gls{si}. 
The variations of the cardiac position were then probed by moving the heart $\pm$\SI{3}{\cm} along \gls{rl}, and \gls{si} axis. 
Translation along the \gls{pa} axis was further restricted to $\pm$\SI{1}{\cm} to prevent any intersections of the heart with the torso. 
Hereon, we will denote the translations as \gls{trsi}, \gls{trrl}, and \gls{trpa}
where a translation is counted positive if it is along an axis as indicated in Figure \ref{fig:transl_rot}).

\emph{Orientation uncertainty} was systematically implemented by rotating the heart around an orthogonal set of eigenaxes derived from the cardiac anatomy.
Eigenaxes were obtained by employing a principal component analysis algorithm on the cardiac mesh, 
and the Gram-Schmidt orthogonalization. 
The first axis, defined as \gls{ab} axis, was determined as the eigenvector associated with the largest eigenvalue obtained with the principal component analysis.
The \gls{rl} axis was computed by applying the Gram-Schmidt orthogonalization to the vector connecting the centers of the left and right blood pools, 
whereas the \gls{pa} axis was determined as the vector orthogonal to the plane spanned by the \gls{ab} and \gls{rl} axes (referred to Fig. \ref{fig:transl_rot}). 
The midpoint of the left ventricular blood pool was considered the origin of this orthogonal system. 
The variations of the cardiac orientation were then explored by rotating the heart by $\pm$\SI{10}{\degree} around the axes \gls{rl}, \gls{pa} and \gls{ab}, respectively (see Fig. \ref{fig:transl_rot}). The rotational transformations were denoted then as 
\gls{rotab}, \gls{rotrl} and \gls{rotpa}, respectively. 

\emph{Uncertainty in size} as induced by imaging uncertainty or breathing, which modulates intra-thoracic pressure and, thus, the filling state and the size of the heart, have been shown to remain under 10\% \cite{holst2018respiratory,claessen2014interaction}. 
The associated variability in size was mimicked by applying $\pm$10\% scaling factors to the heart while keeping its center of mass unaltered.  

We sought to discriminate the contribution of sheer changes in myocardial volume which scales the cardiac sources without affecting the propagation of the depolarization wave fonts, from the changes in the activation and repolarization pattern due to the impact of the myocardial volume size.
We thus investigated \gls{ecg} variability due to heart scaling by prescribing 
i) the activation and repolarization sequence given by the spatially scaled source distribution $V_{\rm m}(\mathbf{\bar{x}},t)$, to mirror that of the reference simulation, and 
ii) the orthotropic conduction velocities associated with the electrophysiological parameter settings 
that yield an altered source distribution, $V^*(\mathbf{x},t)$.
In the latter case, prescribed conduction velocities may slow down or accelerate the epicardial activation, with epicardial breakthroughs being retarded as the myocardial volume and transmural wall width increase, or with precipitated epicardial breakthroughs when the size of the heart is reduced.
Combined with a larger/slower epicardial surface slowing down/hastening total epicardial activation,
the ensuing activation and repolarization pattern will be altered, and, consequently, the ECGs.

\emph{Data analysis} We quantitatively analyzed the variations in the \gls{ecg} depending on the anatomical transformations by comparing their impact upon the \gls{ecg} amplitude on the single R, S, and T peaks for each lead, in terms of both absolute and relative variations. In the former case, we compute:
$$\text{absvar} = |\phi_t^p - \phi_r^p|, $$
while in the latter case, we derive
$$\text{relvar} = \frac{|\phi_t^p - \phi_r^p|}{|\phi_r^p|}, $$
where $\phi_r^p$ and $\phi_t^p$ represent the reference signal and the signal computed with the anatomical transformation $t$, respectively, 
at the instant of each peak $p = R, S$, and  $T$. 
While $\text{absvar}$ expresses the same change in peak amplitude that can be observed by plotting the \gls{ecg}, the relative variation value $\text{relvar}$ points to the geometrical transformations that caused real major variations in the \gls{ecg}, as it identifies these variations relative to the reference scaled amplitude of the \gls{ecg} signal.


\subsection{Uncertainty in electrical conductivities}
For a given distribution of cardiac electric sources, the potential field generated in the torso and the corresponding \gls{ecg} are determined by the tissue-specific electrical conduction inhomogeneities within the torso itself. 
The conductivity within the torso is indeed highly heterogeneous, with substantial differences between compartments (including organs or distinct tissues such as bones, fat, and skin) \cite{bradley2000effects,sanchez2018sensitivity}. Moreover, obtaining precise measures of tissue conduction properties is a difficult endeavor, usually necessitating the use of approximations based on literature values\cite{Keller2010,bear2016optimization}.

To investigate the variability in \gls{ecg} caused by torso heterogeneities, we remove individual tissue-specific conduction inhomogeneities from a reference fully-heterogeneous torso.
Conductive properties of ventricular blood masses, bones, lungs, liver, skin, and subcutaneous fat were taken into account for this study. 
The respective conductivities were taken from \cite{Keller2010,Hasgall2022} and are summarized in Table \ref{Tab:torso_cond}.

\begin{table}[!t]
	\centering 
	\begin{tabular}{ccccccc}
		\toprule
		Blood &Lungs &Bones &Liver &Fat &Skin &Torso tissue\\
		\SI{}{[S/m]} &\SI{}{[S/m]} &\SI{}{[S/m]} &\SI{}{[S/m]} &\SI{}{[S/m]} &\SI{}{[S/m]} &\SI{}{[S/m]} \\
		\hline \\ [-2ex]
		0.66 &0.06 &0.006 &0.35 &0.037 &0.01 & 0.25\\
		\bottomrule
	\end{tabular}
	\caption{Physical conductivities for the considered distinct tissue inside the torso. }
	\label{Tab:torso_cond}
\end{table}	

\emph{Data analysis} The impact of torso conduction inhomogeneities on the amplitude of the \gls{ecg} was quantified by computing the \gls{rmse} \cite{Keller2010,ZAPPON2024112815} for each lead $j$, expressed by:
$$ \text{rmse}_j = \sqrt{\frac{\sum_{i=1}^N(\phi_r^i - \phi_v^i)^2}{\sum_{i=1}^N (\phi_r^i)^2}},$$
where $\phi_r$ and $\phi_v$ represent the  \gls{ecg} signals obtained with reference and varied conductivity settings, respectively, 
and $N$ refers to the number of time samples of the \gls{ecg} signals. 
Variations of \gls{ecg} morphology in each lead, $j$, was instead characterized by the \gls{cc} \cite{Keller2010,ZAPPON2024112815} , computed by:
$$ \text{CC}_j = \frac{1}{s_r s_v} \sum_{i=1}^{N}\left[\phi_r^i - \bar{\phi_r}\right]\left[\phi_v^i - \bar{\phi_v}\right],$$
where $s_r$ and $s_v$ represent the standard deviations of $\phi_r$ and $\phi_v$ over time, 
and $\bar{\phi_r}$ and $\bar{\phi_v}$ are the corresponding arithmetic mean values over time.
Both \gls{rmse} and \gls{cc} were then averaged over all twelve leads.

\subsection{Lead placement uncertainty}
To assess the impact of electrode displacement in the 12-lead \gls{ecg} system, we systematically perturbed the known electrode sites within our calibrated anatomical model. 
Primary tissues such as lungs, atria, blood pools, and general tissue conductivities were used within the model with conductivity values reported in Table \ref{Tab:torso_cond2} as defined as the nominal values within \cite{keller2010ranking}. 

\begin{table}[!b]
	\centering 
	\begin{tabular}{ccccccc}
		\toprule
		Atria &Blood &Lungs &Bones &Fat &Skin &Torso tissue\\
		\SI{}{[S/m]} & \SI{}{[S/m]} &\SI{}{[S/m]} &\SI{}{[S/m]} &\SI{}{[S/m]} &\SI{}{[S/m]} &\SI{}{[S/m]} \\
		\hline \\ [-2ex]
		0.0537 & 0.7 & 0.0389 &0.006 &0.037 &0.01 & 0.22\\
		\bottomrule
	\end{tabular}
	\caption{Physical conductivities for the considered distinct tissues inside the torso in the modeling setup to explore variation in 12 lead ECG electrode placement. }
	\label{Tab:torso_cond2}
\end{table}	

To streamline the automated prescription of electrode positions, we utilized an abstract reference framework defined by \gls{utc}, 
retrofitted to our torso model \cite{gillette2021:_framework}. 
All electrode positions were adjusted from the baseline configuration by $\pm 20$\% along the superior-inferior axis and $\pm 10$\% circumferentially with respect to \gls{utc}. 
Precordial electrodes V1 through V4 were additionally shifted upwards, in the superior direction, 
by $\pm 10$\% of their original position to replicate observed clinical patterns in electrode placement, as reported in \cite{rajaganeshan2008accuracy}. 
For each new position of one of the 9 electrodes, a corresponding new configuration of the 12-lead \gls{ecg} was computed.

T\emph{Data analysis} o quantitatively characterize morphological changes in the 12-lead \gls{ecg} arising from each electrode movement, we computed the time averages of the $L_2$ norm for each lead $j$:
$$ L_2^j = \int_{[t_{init}, t_{init}]} |\phi_r - \phi_v|^2,$$
where $[t_{init}, t_{init}]$ is the time interval over which the \gls{ecg} lead is computed, and $\phi_r$ and $\phi_v$ are the signals obtained with the reference and varied position of the electrodes. 
The resulting values were then averaged across all leads. Furthermore, we normalized the $L_2$ norm to the highest observed $L_2$ values under the given electrode configurations.

\section{Results}

\subsection{Residual Variability}
The relative margin of uncertainty introduced by residual beat-to-beat variability of the ECG was investigated in both healthy subjects and patients treated for \gls{af} and \gls{vt}. 

\begin{figure}[!t]
    \centering
	\includegraphics[width=0.7\textheight]{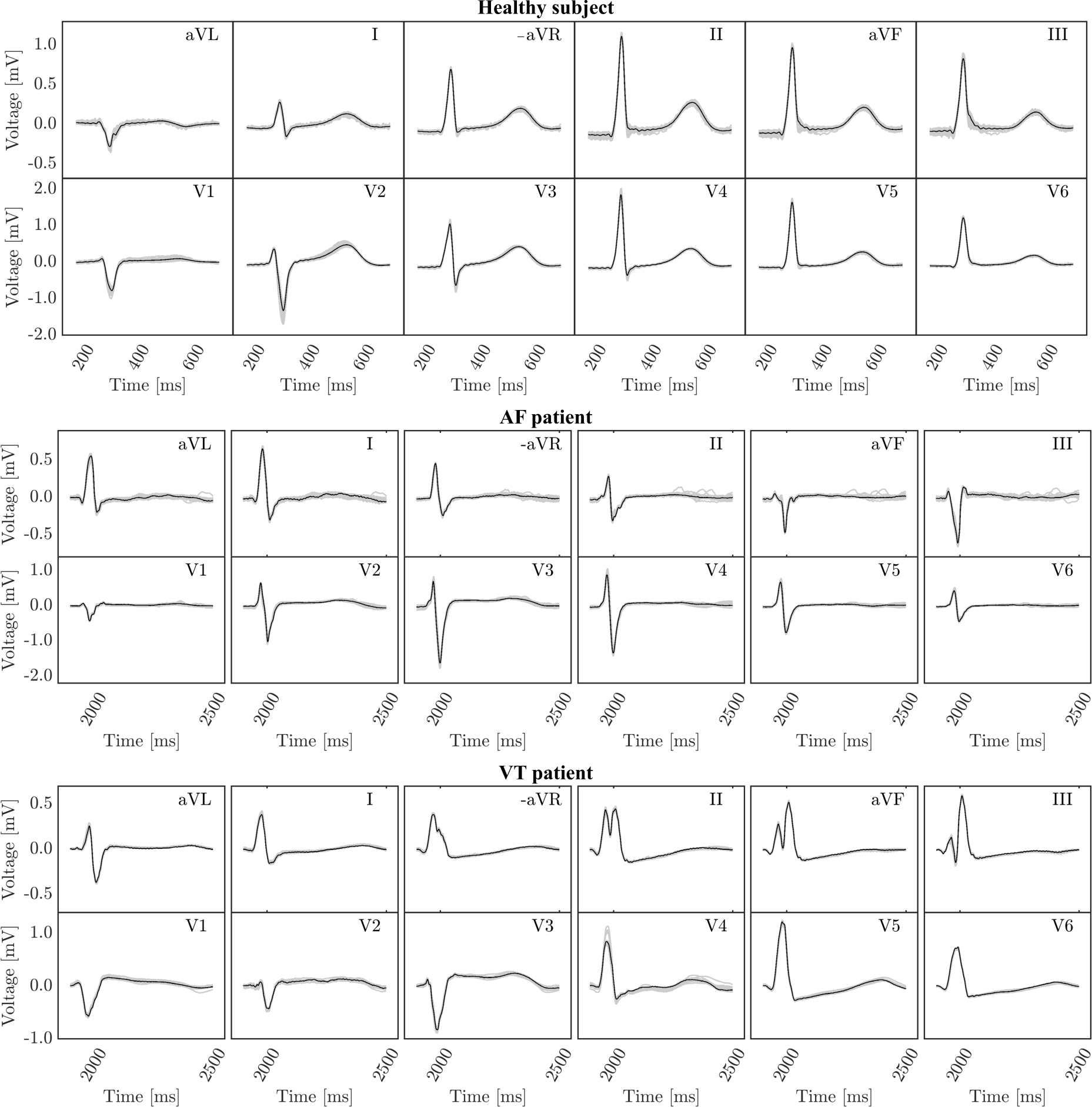}
	\caption{Distribution of the 12-lead \gls{ecg} (grey) and corresponding mean value (black) due to residual beat-to-beat variability of the \gls{ecg} signals. The recorded \gls{ecg}s refer to a representative healthy subject (top), AF patient (middle), and VT patient (bottom).}	
	\label{fig:residual_variability_ecg}
\end{figure}

As expected, in all 14 healthy subjects considered the observed residual beat-to-beat variability in the 12-lead \gls{ecg}s was very minor.
The mean \gls{ecg} beat, representative of a single subject, was plotted relative to the envelope formed by all beats in the beat matrix recorded over a \SI{10}{\second} time window, as shown in Fig.~\ref{fig:residual_variability_ecg}. 
The margin of the envelope in the limb leads was mostly noise-related. Minor variability was seen in the precordial leads V1 and V2 which are located closest to the heart. No noticeable variability was witnessed in leads V3-V6.

Similarly, minor beat-to-beat variability was observed in the majority of \gls{vt} and \gls{af} patients. For such instances, the ECGs were recorded under an intrinsic sinus rhythm over extended acquisition periods lasting at the order of minutes, 
necessary for map construction during the electro-anatomical mapping procedures.
Representative examples are shown in the \gls{vt} and the \gls{af} panels in Fig.~\ref{fig:residual_variability_ecg}. 
Only minor noise-related margins appeared during repolarization in the \gls{af} case, and a minor variability was observed in the R peak of V4 in the \gls{vt} case. 
Overall, in most scenarios residual variability in the \gls{ecg} can be considered negligible,
and, thus, using a median or mean \gls{ecg} as an objective for model calibration
appears suitable and justified.

\subsection{Cardiac anatomical variation - position and orientation}
\begin{figure}[!h]
	\centering
	\includegraphics[width=0.63\textheight]{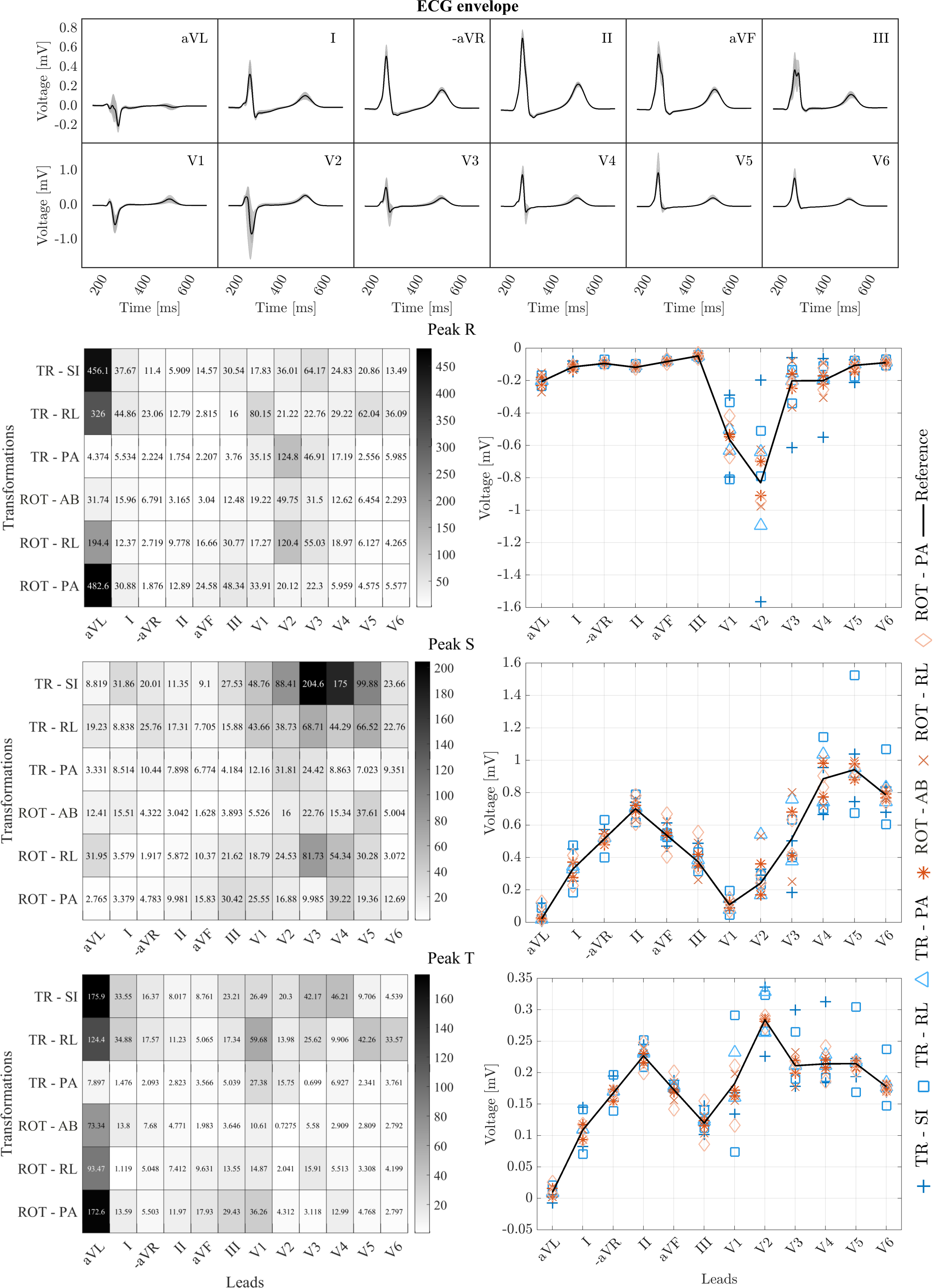}
	\caption{Top: distribution of the 12-lead \gls{ecg} (grey) due to anatomical uncertainties introduced by varying position and orientation of the heart within the torso. The reference \gls{ecg} signal is also depicted (black).
    Bottom-left panel: maximum relative variation (in percentage) of the R, S, and T peak amplitudes caused by each transformation, for each lead. 
    Bottom-right: absolute variation in R, S, and T peak amplitudes caused by each considered transformation, for each lead. 
    }
	\label{fig:heat_comparison}
\end{figure}

The distribution of \glspl{ecg} due to cardiac anatomical variations is qualitatively shown 
as an envelope around the reference \gls{ecg} in Fig.~\ref{fig:heat_comparison}. 
Varying position and orientation of the heart led to 
variations in both \gls{ecg} morphology and peak amplitudes. 
In leads where the lead field axis was closer co-aligned to the maximum dipole, with smaller angle deviations, -- these are lead II or V5 for this vertical-to-normal electrical axis type -- 
morphology was largely unaffected under all transformations, while only minor changes in peak amplitudes were observed.
In leads where the lead field axis was oriented rather orthogonal to the maximum dipole vector, 
such as the limb leads aVL and III, or the precordial leads V1-V4 closest to the cardiac surfaces, larger morphological variations were witnessed.
Specifically, a high relative change in amplitude was observed in aVL, resulting in the appearance of an R wave, and a notable variation in the magnitude and the shape of the S wave. Finally, changes in the T wave were very minor in all leads, as expected.

A quantitative analysis of relative and absolute amplitude variations of each peak across all leads was performed to identify leads where \gls{ecg} morphology is most sensitive to anatomical uncertainty.
As illustrated in Fig. \ref{fig:heat_comparison} (left), a significant change in the amplitude of the R peak relative to baseline values was observed in lead aVL when the heart was translated by \gls{trsi} or \gls{trrl}, or rotated by \gls{rotpa} and \gls{rotrl}. 
Similarly, substantial relative variations of the R peak were noticeable in lead V2 under the transformations \gls{trpa} and \gls{rotrl}.
Regarding the absolute variation in R peak magnitude, the most significant variations occurred in the precordial leads V2, V3, V4, V5, and V6
(see Fig.~\ref{fig:heat_comparison} (second row-right))

The highest changes in peak S were witnessed in the precordial leads V1 to V4. These were predominantly mediated by the transformations \gls{trsi}, \gls{trrl}, and \gls{rotrl}. The largest absolute variation in peak S was induced in lead V2 by the transformations \gls{trsi}
which changed peak S from a maximum negative value of \SI{-1.6}{\milli \volt} down to \SI{-0.2}{\milli \volt} (see Fig.~\ref{fig:heat_comparison} Peak S, right panel).

Finally, absolute variations in the T-wave were minimal (see Fig. \ref{fig:heat_comparison} bottom-right panel). The marked relative variation of the T-wave in aVL is physiologically insignificant as this is caused by the overall very small magnitude of the T-wave in this lead (see Fig. \ref{fig:heat_comparison} bottom-left).

\subsection{Cardiac anatomical variation - size}

\subsubsection{Variable size with prescribed activation sequence}
\begin{figure}[!t]
    \centering
	\includegraphics[width=0.7\textheight]{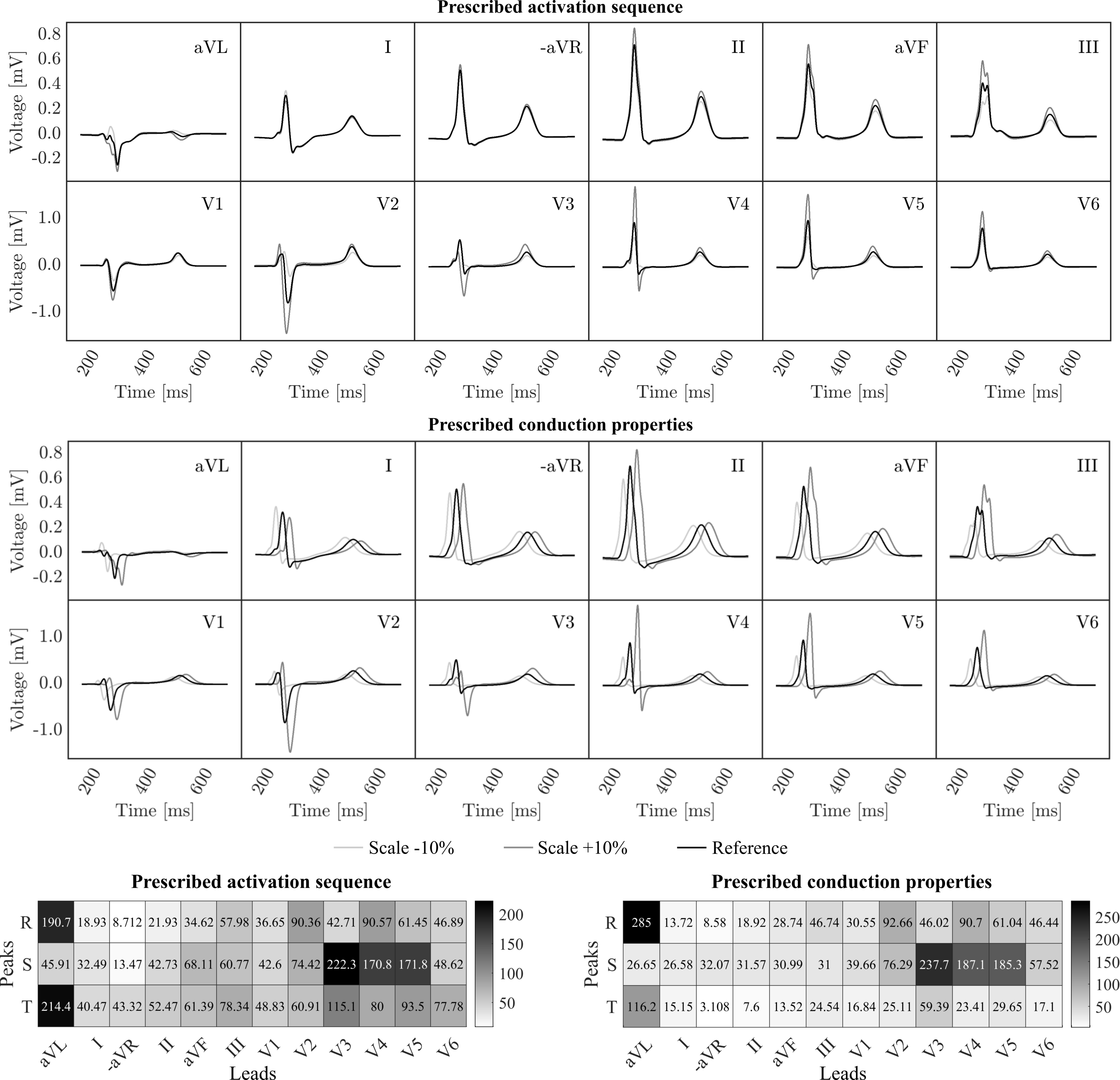}
	\caption{Row 1-2: 12-lead \gls{ecg} obtained by scaling the heart of $\pm$10\% of its original dimension 
     with a prescribed activation sequence (row 1), and with prescribed conduction properties (row 2). Row 3: Maximum relative variation (in percentage) of the R, S, and T peak amplitudes for each lead, due to scaling of the heart when prescribing the activation sequence (left), and when prescribing the conduction properties (right).}
	\label{fig:ecg_scaled}
\end{figure}

The \gls{ecg} obtained by scaling the heart while prescribing the activation sequence as in the reference cardiac anatomy is shown in Fig. \ref{fig:ecg_scaled} (top). 
Prescribing the same activation sequence to a smaller/larger heart is inherently equivalent 
to assume a slower/faster conduction velocity.
As expected, the main effect on the \gls{ecg} was a scaling of signal amplitudes. 
As illustrated in Fig. \ref{fig:ecg_scaled}, a significant change in the amplitude of the S peak relative to the baseline values was especially observed in precordial leads V2-V5.
Specifically, the transition zone of the S wave 
-- in the normal healthy case located between V3-V4 -- was affected. In the smaller heart the transition zone of the S wave shifted towards V2-V3,
whereas in the larger heat a shift towards V4-V5 occurred (see top panel of  Fig.~\ref{fig:ecg_scaled}).
The range of magnitude scaling in the remaining leads was comparable to those observed 
for translation and rotation of the heart.
Also, similar morphological changes such as the appearance of the R wave in lead aVL were witnessed.
However, the overall \gls{ecg} morphology remained largely unaltered, with only minor variations of R and S wave shape and duration in leads V2 and V3, and an altered T wave duration in lead V3. These morphological effects are mediated by the change in distance between cardiac sources and the lead positions of V2 and V3, which are located in the immediate vicinity of the heart.


\subsubsection{Variable size with prescribed conduction properties}
Changes in \gls{ecg} signal magnitudes and morphology obtained by altering the size and prescribing the conduction properties were comparable to those obtained from prescribing the activation sequence (see Fig.~\ref{fig:ecg_scaled}).
However, in contrast, a noticeable time shift of about $\pm$\SI{23}{\milli \second} was witnessed here for the entire trace, mostly corresponding to an earlier/later epicardial breakthrough 
when the cardiac model was scaled to -10\%/+10\% of its dimension, respectively. 

\subsection{Uncertainty in electrical conductivities}
\begin{figure}[!t]
	\centering	\includegraphics[width=0.7\textheight]{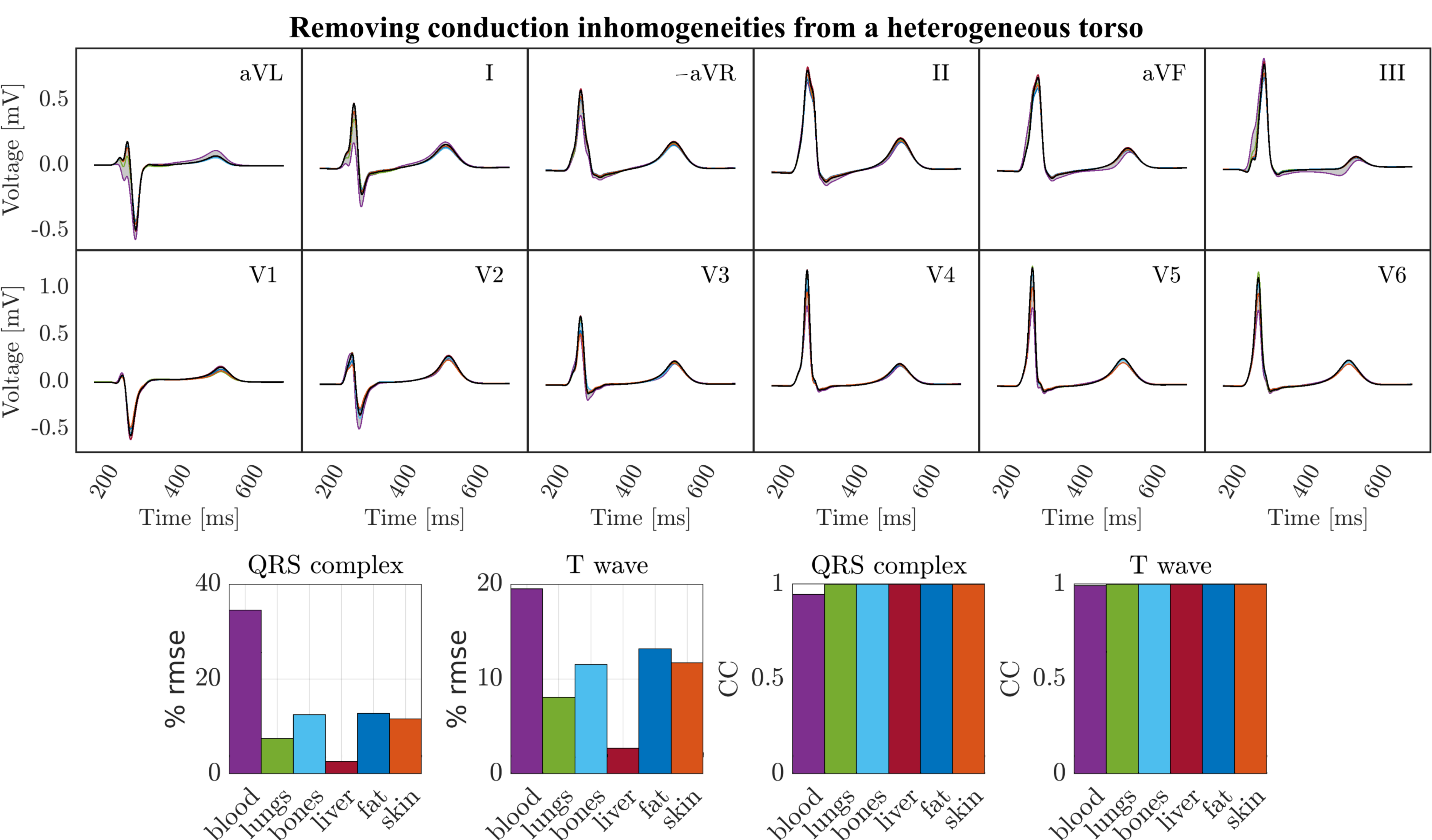}
	\caption{Top: 12-lead \gls{ecg} and corresponding envelopes (depicted in grey) generated by sequentially eliminating organ-specific conduction inhomogeneities from the baseline heterogeneous torso model. The baseline signal (black) referred to the heterogeneous torso. Bottom: the relative \gls{rmse} in percentage (left), and \gls{cc} (right). The colors in the \gls{ecg} representation correspond to the colors of the bar in the \gls{rmse} and \gls{cc} graphs.}	
	\label{fig:ECG_organs}
\end{figure}

The \gls{ecg} envelope resulting from varying the conductive properties of distinct tissues 
within the torso is illustrated in Fig. \ref{fig:ECG_organs}. 
Removing conductive inhomogeneities from the reference heterogeneous torso 
resulted in minor variations in peak amplitudes in all leads, with the exception of the limb leads aVL and I, and the precordial leads V4-V6
where a rather pronounced increase in the R wave amplitude was observed. 
The T wave consistently remained largely unaffected across all leads, with the exception of a very modest increase in the T wave duration 
noticed in the limb leads II, aVF, and III, a slight.

A quantitative assessment of the \gls{ecg} variation due to tissue-specific conductive inhomogeneity
against a fully heterogeneous torso was assessed by computing the \gls{rmse} and the \gls{cc} 
averaged over all leads, as shown in Fig.~\ref{fig:ECG_organs} (bottom panel). 
The blood mass was observed to have a major role in affecting the amplitudes of the QRS complex, 
with a \gls{rmse} of $\approx$37\%, and the T wave with a a \gls{rmse} of $\approx$20\%,
followed by fat, skin, bones and lungs. 
The conductive properties of the liver had a negligible impact upon the ECG.
Finally, the \gls{cc} reported on the bottom-right panel of Fig. \ref{fig:ECG_organs} showed 
a negligible sensitivity of the \gls{ecg} morphology to a variation in conductive properties 
of all tissues, only a minor reduction in \gls{cc} was noticed when the blood mass inhomogeneity was omitted.
 
\subsection{Lead placement uncertainty}
\begin{figure}[!t]
	\centering
 \includegraphics[width=0.70\textheight]{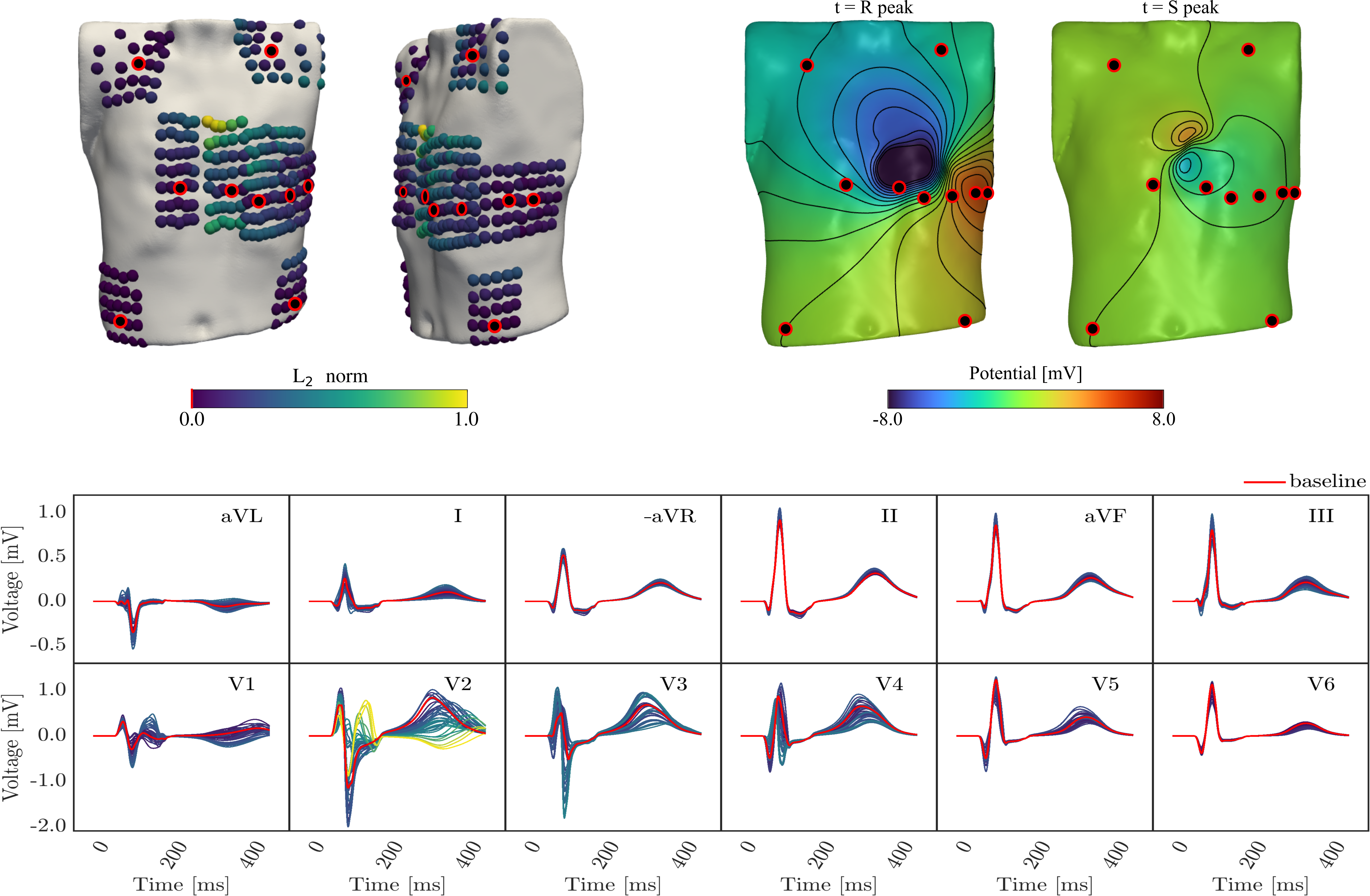}
	\caption{Top-left: Electrode variation on the torso surface from the clinically measured baseline configuration (red and black). Top-right: Body surface potential map at time instances corresponding to the R peak of lead II, and the S peak of lead V4. Bottom: 12 lead \gls{ecg}s under sinus. The coloration of the 12 lead \gls{ecg} and electrode placements corresponds to normalized $L_2$ norm quantifying morphological variation. Baseline (red) 12 lead \gls{ecg} is shown.}	
	\label{fig:electrodes_variability}
\end{figure}

The effect of lead placement uncertainty on \gls{ecg} amplitudes and morphology is illustrated for each electrode placement in the form of the $L_2$ norm computed over all the \gls{ecg} traces, and the corresponding 12-lead ECG (see Fig.~\ref{fig:electrodes_variability}).
Most significant lead placement effects were witnessed in the precordial leads closest to the heart, i.e.\ V1-V4.
Noticeable morphological changes in the \gls{ecg} traces stemmed from a higher placement of the electrodes V1 and V2, 
causing an increasingly apparent RSR pattern and a T-wave inversion in leads V1 and V2,  
and a pronounced S wave in leads V3 and V4.
As expected, limb leads appeared relatively robust against electrode movement. 
This was also the case with the precordial lead V6.   
Finally, slight S-wave slurring and R amplitude elevation were obtained within -aVR and lead II, 
as well as small morphological variations in leads aVL and I.

\section{Discussion}
Advanced \gls{cdt} generation and calibration methods have been developed for finding model parameters that minimize the discrepancy between simulated and measured \gls{ecg} \cite{grandits2023digital,gillette2021:_framework,CAMPS2024103108}.
While feasible, in principle, the fundamental issue of non-uniqueness of the solution persists, 
that is, more than one parameter set may exist that calibrates the model equally well 
to the observed \gls{ecg}. 
Moreover, due to the nature of the clinical data acquisition process and the computational workflows for the generation of \gls{cdt} models, inconsistencies between real and virtual \gls{cdt} replicas of anatomy and physics in a patient, inevitably arise. These lead to discrepancies between real and simulated \gls{ep}, although the observations used for calibration -- the \gls{ecg} -- might be very similar or identical.

In our study, we investigate the impact of the major sources of inconsistencies upon the simulated \gls{ecg}.
These comprise
i) the residual beat-to-beat variability reflected in morphological alterations of the \gls{ecg}
ii) observational uncertainties due to technical limitations 
in anatomical imaging and \gls{ecg} recording, impeding an accurate synchronous measurement of all model parameters contributing to the genesis of the \gls{ecg} \cite{keall2006management,hawkes2005tissue}. 
The latter includes the generation of computational cardiac and torso anatomies from tomographic images, which are modulated by the subject/patient breathing, the inability to measure electrical model parameters, such as the heterogeneity of electrical conductivities throughout the torso, and variability in the lead placement.
If the uncertainty in the observed ECG is high, 
and the \gls{ecg} predicted by the computational representation is overly sensitive to these inaccuracies in anatomical and electrical parameters, the calibration of a model based on a measured \gls{ecg} is effectively impeded, 
or may be unreliable.

Here, we aim to provide a general view of the impact of these uncertainties on the simulated \gls{ecg}, and to investigate to which extent they may affect the ability to calibrate a computational model of cardiac \gls{ep}.
Our qualitative analysis of residual variability in both healthy subjects and \gls{af} and \gls{vt} patients showed a rather tight envelope in the \gls{ecg} waveforms, 
indicating that using an arbitrarily chosen \gls{ecg} beat, or a mean \gls{ecg} beat, is admissible, and does not introduce any significant uncertainty.
Further, the effects on \gls{ecg} morphology and derived diagnostic markers of uncertainties in location, orientation, and size of the heart, in electrical torso conductivities, and lead placement, were computed. We moreover considered a worst-case scenario, defining uncertainty outputs falling outside the level of accuracy obtainable with careful dedicated clinical data acquisition.
While all these factors were shown to impact the \gls{ecg} waveforms, they primarily contributed to generating \gls{ecg} traces that retained the most important morphological features and diagnostic markers, and closely adhered to the ground truth \gls{ecg}.

Overall, our results suggest that an accurate calibration of a \gls{cdt} is not impeded by observational uncertainties and model inconsistencies.
Whether a set of parameters that optimally calibrates a model to the \gls{ecg} is uniquely identifiable remains an open question, beyond the scope of our study.

\subsection{Residual variability}
The calibration of a \gls{cdt} to replicate the patient heart's intrinsic activation and repolarization pattern
based on the \gls{ecg}, relies upon a stable rhythm.
However, beat-to-beat alterations cause residual variability in repetitive measurements of the \gls{ecg},
leading to an ensemble of \gls{ecg} waveforms, and the problem of selecting a representative beat in the \gls{ecg} 
to be used for calibration.

Beat-to-beat variability is a well-known and extensively studied phenomenon.
In healthy subjects, the heart rate is known to be always variable, even under resting conditions \cite{hasan2012relation,yeragani2000effect,APPEL19891139,vsipinkova1997effect}.
However, heart rate variability \emph{per se} is not a relevant factor as only the temporal onset of the heart beat is modulated, but the electrical activation and repolarization sequence remains largely unaffected, that is,
the \gls{ecg} waveform does not change.
With respect to \gls{cdt} calibration, only beat-to-beat alterations that affect the \gls{ecg} waveform noticeably enough to indicate a change in the cardiac activation or repolarization pattern, are considered of interest.
Such alterations in morphology, including very small amplitude variations at a \SI{}{\micro \volt} scale \cite{eckberg2003topical,FRLJAK2003267,APPEL19891139,atiga1998beat,pueyo2016experimentally,APPEL19891139,huikuri1996abnormalities, konta1990significance,kay1983torsade}, are often utilized as a clinical biomarker,
indicative of a propensity for arrhythmias to occur. 

In our study, we investigate beat-to-beat \gls{ecg} morphological variability in cohorts of healthy subjects, \gls{af} and \gls{vt} patients.
In the healthy subjects, the \gls{ecg} envelope caused by the residual variability was narrow, suggesting that such variability could be negligible for an \gls{ecg} based calibration of a \gls{cdt}.

This finding also held true for the patient cohorts treated with \gls{af} or \gls{vt} ablation therapy. Our analysis of \glspl{ecg} recorded during interventions in 36 \gls{af} and 17 \gls{vt} patients showed very limited variability in most cases. This was expected, as \glspl{ecg} were recorded during mapping studies where stable activation patterns are required for map construction. In cases where more noticeable changes in \gls{ecg} morphology occurred, clustering of \gls{ecg} morphologies would be necessary, and \glspl{cdt} would need to be re-calibrated for each identified cluster. As high-fidelity \gls{ecg} calibration becomes increasingly feasible and streamlined \cite{grandits2023digital}, facilitating calibration to an ensemble of \glspl{ecg}, the identified model parameters leading to a good match may offer insights into the underlying physiological mechanisms.

\subsection{Cardiac anatomical variation}
Clinical ECG recording and anatomical image acquisition are not synchronous. Therefore, the precise anatomical configuration of the heart at the time of \gls{ecg} recording is generally unknown. Moreover, a 3D whole heart anatomical MRI scan, as used in this study, is typically acquired during diastasis, as this phase of the cardiac cycle provides the longest window with minimal cardiac motion. However, during diastasis, the ventricles are smaller than their end-diastolic configuration, when electrical depolarization associated with the QRS complex occurs, and larger than their end-systolic configuration, when repolarization associated with the T-wave takes place \cite{klein2017cardiac,budoff2009determination}. Other anatomical and technical factors during image acquisition are moreover responsible for the increase of uncertainty in the cardiac anatomy reconstruction, and its relation to the recorded \gls{ecg}.

For instance, discrepancies between cardiac anatomy in a patient and its representation in an anatomical model inevitably arise due to a combination of physiological and technical factors. Physiologically, the position, orientation, size, and shape of the heart are highly variable, influenced by factors such as breathing \cite{holst2018respiratory,claessen2014interaction,Shechter2004,lendrum1979respiratory,JAGSI2007253,Shechter2004}, body posture \cite{rodeheffer1986postural,rapaport1966effect}, as well as cardiac motion and deformation over a heartbeat itself \cite{ashikaga2008changes,carlsson2004total}. Breathing alters the relative location and orientation of the heart within the torso, while the size of the heart is specifically affected by maneuvers during acquisition, such as a breath hold, which changes the intra-thoracic pressure and, thus, the blood volume in the cavities \cite{noseworthy2005impact,sakuma2001effect}.
Moreover, prolonged time intervals often occur between imaging and \gls{ecg} recording. This can result in significant changes in the heart's size due to variations in heart rate or the patient's fluid status.

Technical factors include limited spatial resolution and contrast in the images, hindering the accurate segmentation of all relevant structures. Additional challenges include the displacement between slices in sequential 2D image acquisitions, which are temporally registered using a navigator or \gls{ecg} gating, the registration errors between the heart and the whole torso imaging, and the device artifacts that partially obscure the cardiac anatomy \cite{counseller2023recent,kalisz2016artifacts,scott2009motion,gamper2007diffusion}. 

All these factors combined contribute to cardiac anatomical uncertainty, affecting the heart's shape, size, and position not only anatomically but also as an electrical source relative to the \gls{ecg} recording site. This uncertainty effectively alters the \gls{ecg} waveform.

The influence of the heart's position and orientation on the \gls{ecg} has been explored in vivo and with real patients in several studies. In \cite{hoekema2001geometrical,van2000geometrical}, the authors estimate the heart's position relative to the torsos of 25 subjects by comparing body surface potential maps with the cardiac and torso geometry obtained from MRI scans, and then compute the corresponding \gls{ecg} through computational models. In \cite{hoekema2001geometrical}, the pericardial position of each subject is inferred through an electrocardiographic inverse solution. These 25 solutions are then used to compute the statistical variability caused by geometric error on the \gls{ecg}. However, the authors are unable to separate the influence of physiological and geometric variability on the \gls{ecg}.
A direct measurement of the heart's position compared to the torso shape and the position of the lead electrodes is performed in \cite{van2000geometrical}, where the authors used MRI images of both the heart and the torso to create 3D triangular meshes. The solid angle of the heart relative to the torso is then computed, along with other thoracic indices, and these are accounted for in the analysis of \gls{ecg} variability. However, this study does not isolate the effect of cardiac size and position variability within a single subject, but rather represents a population study. Single-subject \gls{ecg} variability due to heart position is addressed in \cite{MACLEOD2000229}, where the authors analyze \gls{ecg} changes due to heart position using an instrumented, isolated heart suspended in a torso-shaped electrolytic tank. Changes in heart position are accounted for by moving the heart in a three-axis orthogonal system built into the torso, to match uncertainty in cardiac position due to the MRI imaging process of human subjects. While the applied cardiac transformations resemble our heart translations in both direction and magnitude, the hearts used are from animals, the torso is not realistic, and the \gls{ecg}s and their variations are not clearly represented.


A more in-depth analysis of \gls{ecg} variations \emph{in silico} has been previously investigated 
in a small number of studies. 
In \cite{nguyen2015silico}, the authors exploited the effect of the cardiac position 
on pathological QRS complex in five patients. 
The simulated \gls{ecg}s of the reference patients were initially validated against clinical data and then compared with the same \gls{ecg} signals obtained by displacing the cardiac domain using upper-lower and right-left rigid translations and horizontal-vertical rigid rotations. Variability of the QRS complex due to heart-torso geometrical uncertainty was also analyzed in \cite{minchole2019mri}. In this study, starting with 5 cardiac and 5 torso geometries, a cohort of 625 heart-torso anatomical models was created by combining the cardiac and torso models, and defining two rigid translational directions and two rigid rotational directions for the heart. The \gls{ecg}s were compared, and a qualitative assessment of the reliability of the computed \gls{ecg}s against clinical data was conducted. However, in combining the heart and torso geometries, additional factors such as electrophysiological activation and variations in cardiac shape between anatomies contribute to \gls{ecg} generation. Moreover, both studies focus solely on the QRS complex and consider a limited number of translations and rotations. Additionally, the heart's size is not included as a factor affecting the \gls{ecg}, and the set of simulated \gls{ecg}s are compared against a single reference signal for each patient, without considering beat-to-beat variability.

Our work aims to model all possible transformations of the heart based on \emph{in vivo} analysis, including heart scaling, by exploring the limit values of these transformations in a single patient. This approach effectively defines the range of variations in the \gls{ecg} signals without introducing additional factors. Additionally, we extend the analysis to the T wave and compare the results with the beat-to-beat variability of the reference \gls{ecg}. Our results showed that \gls{ecg} shape variations are overall minor. Furthermore, we observed different sensitivities of the R, S, and T peaks to cardiac geometrical transformations in different leads, with aVL, V3, and V4 being the most affected. The most influential transformations were translations in the superior-inferior direction, rotations in the postero-anterior direction, and heart scaling.

We highlight the potential for increasing and decreasing depolarization and repolarization duration due to changes in heart size when prescribing conduction parameters, as opposed to a time shift of the \gls{ecg} signals but with the same wave duration when prescribing the activation sequence. This underscores the prominent role of cardiac activation in \gls{ecg} generation.

Overall, our results suggest that the dimension, position, and orientation of the heart have a limited influence on the calibration of \gls{cdt}s.

\subsection{Uncertainty in conductive properties}
"While efficient, anatomically detailed segmentation of tissues and organs of the torso from high-quality clinical images is becoming increasingly feasible \cite{wasserthal2023:_totalsegmentator}, measuring their respective electrical properties is challenging and infrequently undertaken. Consequently, modeling studies often rely on values reported in the literature \cite{gillette2021:_framework,Keller2010} or public databases, with potential \emph{ad-hoc} adjustments to better fit the \gls{ecg} signal \cite{bear2016optimization}.

Computational analysis of the effect of tissue-specific conduction properties within the torso on the \gls{ecg} has been conducted previously 
in \cite{bradley2000effects, keller2010ranking}. 
In \cite{bradley2000effects} the effect of each of the lungs, skeletal muscle, and subcutaneous fat inhomogeneities was investigated in a single patient model. While quantitative indices were taken into consideration, a highly simplistic lumped dipole model was used, and the effect of low skin conductivity was not considered. 
Similarly, in \cite{Keller2010}, the effect of major tissue-specific inhomogeneities on the \gls{ecg} was comprehensively analyzed using a simplified model of the cardiac sources. These sources were represented by a monodomain model in the ventricles and a cellular automaton in the atria, interpolated onto a background torso mesh, and used as a volumetric source for solving the forward problem. Consistent with our study, the conductivity of the blood pool was identified as the most significant factor. Keller et al. also highlighted the importance of fat, ranking it closely behind blood conductivity, and the negligible effect of the liver. However, unlike our findings, they observed that lungs exerted a greater influence than bones. 
Nevertheless, a limitation of both studies is the lack of a reference clinical \gls{ecg} for comparison with the simulated results and the absence of simulated clinical 12-lead \gls{ecg}s. Without these, an analysis of morphological effects and their diagnostic interpretation is precluded, and the assessment of the validity and fidelity of the source model is limited.

Overall, our findings indicate a rather marginal impact of conductive inhomogeneities on the 12-lead \gls{ecg} (see Fig.~\ref{fig:ECG_organs}, suggesting that their consideration -- with the exception of the blood pool conductivity -- is not essential for the calibration of \glspl{cdt} to the \gls{ecg}. 

\subsection{Lead placement uncertainty}
In clinical practice, variations in the morphological patterns of recorded 12-lead \gls{ecg} systems can result from electrode positioning discrepancies, a phenomenon noted even among trained practitioners \cite{rajaganeshan2008accuracy,schijvenaars2008intraindividual}. Wenger et al. \cite{wenger1996variability} demonstrated that precordial electrodes are typically positioned approximately \SI{2.9}{\cm} away from the standard clinical placements, with deviations exceeding \SI{1.6}{\cm} observed upwards in 50\% of V1 and V2 placements, and leftward and downward in up to 50\% of V4 and V6 placements. Recent clinical studies have sought to assess the impact of electrode positioning by manually adjusting electrode positions or selecting specific subsets from \gls{bspm} recordings \cite{medani2018accuracy,kania2014effect}. Nevertheless, these studies often encounter challenges such as low or inconsistent spatial resolution due to manual placement or reliance on artificial signal interpolation from \gls{bspm} recordings. Additionally, the accuracy of information regarding the actual underlying cardiac electrical activity is frequently limited

An in silico study investigating the effect of electrode displacement on five patients was conducted in \cite{nguyen2015silico}. While this study provided a consistent experimental setup with precise electrode placement and simulated physiological conditions, its probe is restricted to displacing solely the precordial electrodes in downward directions, therefore not exploring the entirety of potential electrode sites.

Our study aims to explore a consistent part of the thoracic surface, by moving the electrodes on the longitudinal axis of the torso, both upwards and downwards, and circumferentially with respect to the \glspl{utc}, thus covering right and left sites, including the average \SI{2.9}{\cm} displacement found in \cite{wenger1996variability}. 
Furthermore, all electrodes were displaced, including the limb leads. Our findings indicate that electrode displacement primarily affects the R wave, particularly in the precordial leads, as well as in leads I and III. While the \gls{ecg} morphology remains largely intact across most leads, we observed slight to more pronounced variations in amplitude in the limb leads.
Among the precordial leads, V2 is notably sensitive to upward and downward shifts of its corresponding electrodes, particularly affecting the amplitude of the R wave. However, significant changes in the waveform, such as the appearance of an RSR pattern and inverted T waves, are only observed with an extreme displacement of the V2 electrode. When electrodes are moderately displaced, variations in the \gls{ecg} resemble those induced by shifting the heart in \gls{trsi} directions, as reported in \cite{nguyen2015silico}. Our findings affirm that the distance between the heart and electrode placement, especially for precordial leads closest to cardiac sources, is crucial in determining \gls{ecg} morphology.

With respect to the \gls{cdt}s, we demonstrated that electrode displacement should be considered but does not fundamentally define the \gls{ecg} morphology or overall calibration.

\section{Limitations}
"A major limitation of this study is the use of a single-patient anatomical model. Given that the shape of the heart, the cardiac conduction system, and the torso are crucial factors influencing the spatio-temporal evolution of cardiac sources and their manifestation as torso potentials, some aspects of our findings may not be generalizable.
Specifically, the modeled subject had a vertical electrical axis, suggesting that changes in \gls{ecg} morphology may not manifest similarly in other subjects. Nevertheless, our overall observation regarding the minimal impact of model inconsistencies on \gls{ecg} morphology should remain applicable, regardless of the electrical axis type.


While the selected set of cardiac anatomical transformations covers the complete range of possible anatomical variations of the heart within a single patient, our experiments only consider individual transformations. The potential effects on the \gls{ecg} resulting from combined anatomical variations are not explored. However, considering the impact of anatomical variations, which result in a relatively narrow distribution, it seems unlikely that combined anatomical transformations would significantly widen the range of \gls{ecg} variations.

In examining the \gls{ecg} dependence on electrical tissue properties, we tested only a single conductivity value for each organ,  although measurements suggest a certain variability. Additionally, the conductivities of the heart and their anisotropy ratios were held constant, despite their known high uncertainty \cite{roth1997:_conductivities}, which can significantly influence \gls{ecg} morphology \cite{ushenin2021parameter,Sebastian2008}.


\section{Conclusions}
Our results indicate that uncertainties related to anatomical position, orientation, size, electrical conductivity heterogeneity of tissues and organs in the torso, as well as lead placement, have limited impact on \gls{ecg} morphology and diagnostically relevant \gls{ecg} features. This finding aligns with the narrow distribution of \Gls{ecg} due to residual beat-to-beat variability observed in both healthy subjects and patients.
Our findings suggest that the \gls{ecg} morphology is robustly defined by the electrical activation and repolarization pattern, and, to a much lesser extent, by the considered inconsistencies.  Consequently, these factors do not significantly impede \gls{ecg}-based \gls{cdt} calibration. However, the core challenge of calibration 
-- the unique identification of model parameters --
remains a challenge, although it was beyond the scope of our study.

\section*{Acknowledgments}

        This research was supported by grants from the Austrian Science Fund (FWF) grants no. I6467-B to G.P. and  10.55776/ESP592 to K.G., the European Union’s Horizon 2020 research and innovation program under the Marie Sk\l{}dowska-Curie grant TwinCare-AF  agreement no. 101148636 to E.Z., 
        and in part the Austrian Research Promotion Agency (FFG) grant FO999891133 to M.A.F.G.
        E. Z. acknowledges her membership to INdAM GNCS - Gruppo Nazionale per il Calcolo Scientifico (National Group for Scientific Computing, Italy).
        \begin{figure}[!h]
        \centering
        \includegraphics[width = 0.4\textwidth]{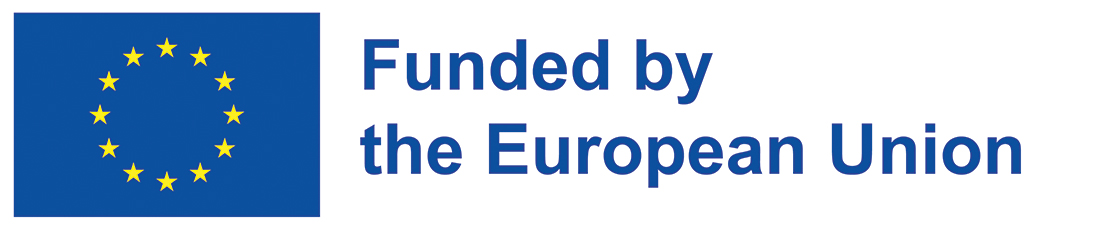}
        \end{figure}
        
\bibliography{Ref}

%
%

\end{document}